\documentclass[a4paper,12pt]{article}

\usepackage{citesort}
\usepackage{epsfig}

\makeatletter

\@addtoreset{equation}{section}
\makeatother
\def\lsim{\;\raise0.3ex\hbox{$<$\kern-0.75em\raise-1.1ex\hbox{$\sim$}}\;}
\def\gsim{\;\raise0.3ex\hbox{$>$\kern-0.75em\raise-1.1ex\hbox{$\sim$}}\;}
\def\beq{\begin{equation}}   \def\eeq{\end{equation}}
\def\bea{\begin{eqnarray}}   \def\eea{\end{eqnarray}}
\def\ba{\begin{array}}   \def\ea{\end{array}}

\def\noi{\noindent}

\def\l{\lambda}
\def\k{\kappa}
\def\tanb{\tan\!\beta}
\def\NMHDECAY{{\sf NMHDECAY}}

\begin{document}

\centerline{\Large\bf The upper bound on the lightest Higgs mass}
\vskip 3 truemm
\centerline{\Large\bf in the NMSSM revisited}
\vskip 2 truecm

\begin{center}
{\bf Ulrich Ellwanger}\\
\vskip 5 truemm
Laboratoire de Physique Th\'eorique\footnote{Unit\'e mixte de 
Recherche -- CNRS -- UMR 8627} \\
Universit\'e de Paris XI, F-91405 Orsay Cedex, France\\
\vskip 5 truemm
{\bf Cyril Hugonie}\\
\vskip 5 truemm
Laboratoire Physique Th\'eorique et Astroparticules\footnote{Unit\'e mixte de 
Recherche -- CNRS -- UMR 5207}\\
Universit\'e de Montpellier II, F-34095 Montpellier, France
\vskip 2 truecm

\end{center}

\begin{abstract}
We update the upper bound on the lightest CP even Higgs mass in the
NMSSM, which is given as a function of $\tanb$ and $\l$. We include the
available one and two loop corrections to the NMSSM Higgs masses, and
constraints from the absence of Landau singularities below the GUT
scale as well as from the stability of the NMSSM Higgs potential. For
$m_{top}$ varying between 171.4 and 178~GeV, squark masses of 1~TeV and
maximal mixing the upper bound is assumed near $\tanb\sim 2$ and varies
between $139.9$ and $141.4$~GeV. \end{abstract}

\vskip 1 truecm

PAC numbers: 12.60.Jv, 14.80.Cp, 14.80.Ly

\vfill
\noi December 2006\\
\noi LPT Orsay 06-77\\
\noi LPTA Montpellier 06-62

\newpage 
\pagestyle{plain} 
\baselineskip 18pt

\section{Introduction} 

Supersymmetric extensions of the standard model predict quite generally
at least one relatively light Higgs boson. Hence, as soon as results
from future collider experiments provide us with informations on the
mass of at least one Higgs boson, we will be able to put constraints on
possible supersymmetric extensions of the standard model.

To this end we need to know, as accurately as possible, how the Higgs
boson masses depend on the nature of the supersymmetric extensions of 
the standard model, and on the parameters of these models. (We hope, of
course, to get independent informations on these parameters from direct
sparticle detections in the future.)

In the MSSM, corresponding calculations have been pushed to a fairly
high accuracy, including many two-loop corrections. Recent reviews on
the lightest Higgs boson mass in the MSSM can be found in
refs.~\cite{r1,r2,r3}.

In the present paper we discuss the simplest version of the
NMSSM~\cite{nmssm} with a scale invariant superpotential
\beq\label{1.1e}
W = \l  {S}  {H}_u  {H}_d + \frac{\k}{3} \,  {S}^3 + \dots\ , 
\eeq
which is the only supersymmetric extension of the Standard Model where
the weak scale originates from the soft susy breaking scale only, i.e.
where no supersymmetric dimensionful parameters as $\mu$ are present in
the superpotential. 

It is well known~\cite{nmssm} that the lightest Higgs boson in the
NMSSM can be heavier than the one of the MSSM due to additional terms
in the tree level Higgs potential proportional to~$\l^2$; the
additional contribution is 
\beq\label{1.2e}
\Delta m_h^2 = \frac{\l^2}{g^2} M_Z^2 \sin^2 2\beta \ .
\eeq 
If one requires the absence of a Landau
singularity for $\l$ below the GUT scale, $\l$ is bounded by
$\sim 0.7$ from above ~\cite{nmssm}, leading still to an upper bound on
the mass of the lightest Higgs boson that is, however, larger than in
the MSSM. Thus, future measurements of the Higgs boson mass could serve
to distinguish these two models, provided that we know the difference
between the upper bound on the mass of the lightest Higgs boson in the
different models.

At present, the radiative corrections in the NMSSM have not been
computed to quite the same accuracy as in the MSSM. Of course,
radiative  corrections in the NMSSM that are proportional to the
quark/lepton Yukawa couplings and the gauge couplings only are the same
as in the MSSM, but there are many additional contributions involving
the new Yukawa couplings $\l$ and $\k$ in the superpotential
in eq.~(\ref{1.1e}), and the associated soft trilinear couplings
$A_\l$ and $A_\k$.

The one loop corrections in the NMSSM induced by t and b quark/squark
loops have been computed already some time ago~\cite{rad1}, and the
dominant two loop corrections ($\sim h_t^6$ and $\sim h_t^4 \alpha_s$),
that are the same as in the MSSM, have been included in an analysis of
the NMSSM Higgs sector in ref.~\cite{rad2,rad2b}.

The leading logarithmic one loop corrections to the lightest Higgs mass
in the NMSSM proportional to the electroweak gauge couplings $g$ or
NMSSM specific Yukawa couplings $\l$, $\k$ ($\sim g^4,
g^2\l^2, g^2\k^2, \l^4, \k^4$) have been computed
only recently ~\cite{rad3}. They are included in the latest version of
the code \NMHDECAY~\cite{nmhdecay,nmhdecay2,nmhdecay3},
where the NMSSM Higgs masses, couplings and
branching ratios are computed as functions of the parameters in the
Lagrangian of the model.

This code checks also the absence of a Landau singularity for $\l$
below the GUT scale using the two loop renormalization group equations,
and susy threshold effects around the susy scale. This procedure is
numerically relevant, since $\Delta m_h^2$ depends on $\l$, and
the upper bound on $\l$ depends on $\tanb$ (via the top
Yukawa coupling $h_t$) and $\k$.

It is the purpose of the present paper to review the upper bound on the
lightest Higgs boson mass in the NMSSM, using the up-to-date knowledge
of the corresponding radiative corrections.

Instead of investigating far-fetched regions in parameter
space that serve to obtain very conservative bounds, we proceed as
follows: For the soft terms that are relevant for the sparticle
spectrum, we chose universal squark and slepton masses of $1$~TeV, and
trilinear couplings of $2.5$~TeV (that practically maximize the one loop
radiative corrections to the Higgs boson mass). For the gaugino masses
we take $M_1 = 150$~GeV, $M_2 = 300$~GeV and $M_3 = 1$~TeV in rough
agreement with universal gaugino masses at the GUT scale. For the top
quark pole mass we present results both for $m_{top} = 171.4$~GeV (the
latest central value obtained by the Tevatron Electroweak Working Group
\cite{teva}) and a very conservative upper limit of $m_{top} = 178$~GeV.
The NMSSM specific Yukawa couplings and trilinear soft terms $\l$, $\k$, $A_{\l}$
and $A_{\k}$ as well as the effective $\mu$ parameter ($\mu = \l s$ in the NMSSM)
are chosen such that the lightest Higgs
boson mass is maximized, without violating constraints from
perturbativity of the Yukawa couplings at the GUT scale, nor
phenomenological constraints on CP odd or charged Higgs masses and
couplings. To this end a numerical analysis is required, that is
performed using the updated version of \NMHDECAY~\cite{nmhdecay3}.
The upper bound on the lightest Higgs mass is then given as a function
of $\tanb$ and $\l$.

For the same choice of the above soft terms, we present the upper
bound  on the lightest Higgs mass in the MSSM limit $\l\to 0$
as obtained with \NMHDECAY. This result can be
compared to values obtained from analytical or numerical analyses in the
MSSM, that include radiative corrections that are still absent in
\NMHDECAY: these are notably electroweak one-loop corrections $\sim
g^4$ beyond the LLA, and non-dominant two-loop corrections (involving
less than two powers of large logarithms) $\sim h_t^6$ and $\sim h_t^4\
\alpha_s$ beyond the ones that follow from an RG improvement of the one
loop corrections~\cite{2lrgemssm} (which are included).

For soft terms  as above, $m_{top} = 178$~GeV, $M_A = \mu = 1$~TeV and
$\tanb = 10\ $  {SuSpect} gives $m_h \sim 128.5$~GeV (taken from
ref.~\cite{r2}), FeynHiggs $m_h \sim 134$~GeV (taken from ref.~\cite{r3}),
and \NMHDECAY\ $m_h \sim 128.6$~GeV. This allows to estimate the
uncertainties on $m_h$ due to the radiative corrections not included in
\NMHDECAY, following the discussions in~\cite{r2,r3} that we will
not repeat here.

The striking effect in the NMSSM is that the maximal value of $m_h$ is
not assumed for large $\tanb$ as in the MSSM, but at low
$\tanb \sim 2$ due to the tree level term noted above. There we obtain
$m_h \sim 139.9$~GeV for $m_{top} = 171.4$~GeV, and $m_h \sim 
141.4$~GeV for $m_{top} = 178$~GeV (for the same other parameters as 
above). For larger values of $\tanb$ the upper bound on $m_h$ decreases
in the NMSSM. For $\tanb \gsim 10$  it hardly exceeds the MSSM value
given above, since the effect of the tree level term becomes small. 
For small $\tanb \lsim 2$ the absence of a Landau singularity below
$M_{GUT}$ restricts $\l$ more strongly from above, due to the large top
Yukawa coupling $h_t$. This implies that present lower limits on $m_h$
from LEP still lead to a lower bound on $\tanb$ of $\sim 1.3$ in the
NMSSM.

It must be noted that in particular regions of the parameter space of
the NMSSM the upper bound on $m_h$ discussed here can be misleading: 

In principle, a singlet-like CP even Higgs boson can be lighter than
the lightest doublet-like CP even Higgs boson (with non-vanishing
couplings to the $Z$ boson) in the NMSSM. Strictly speaking, the upper
bound on the lightest CP even Higgs boson discussed here is then still
valid.

However, a singlet-like CP even Higgs boson would have been practically
undetectable at LEP due to its vanishing coupling to the $Z$ boson.
Fortunately, if the lightest CP even Higgs boson is a pure singlet in
the NMSSM, the upper bound on $m_h$ discussed here applies then to the
lightest doublet-like CP even Higgs boson. On the other hand, if the
lightest CP even Higgs boson is only approximately a singlet, the
lightest doublet-like CP even Higgs boson can be heavier than the upper
bound on $m_h$ discussed here. 

A similar reasonning applies to the situation where the doublet-like CP
even (SM like) Higgs boson decays into singlet like (mostly two CP odd)
scalars~\cite{lighta}. Then, the detection of the SM like Higgs boson can be
very challenging, even if its mass satisfies the upper bounds discussed here.
Hence, although the upper bound on $m_h$ presented here is always valid,
it may refer to a state that is difficult to detect.

\section{The upper bound on the lightest Higgs Boson
mass in the NMSSM}

In order to find the regions in the parameter space of the NMSSM that
maximize the upper bound on the lightest CP even Higgs Boson mass, it
is helpful to take a look at the CP even Higgs mass matrix at tree
level. In the basis $(H_{u}, H_{d}, S)$ and using the
minimization equations in order to eliminate the soft masses
squared, it reads:
\beq\label{2.1e}
{\cal M}_S^2  = \left[
\ba{ccc}
g^2 h_u^2 + \mu \displaystyle\frac{h_d}{h_u}\, (A_\l + \nu) &
(2\l^2 - g^2) h_u h_d - \mu (A_\l + \nu) &
2\l h_u \mu - \l h_d (A_\l + 2\nu) \\ &
g^2 h_d^2 + \mu \displaystyle\frac{h_u}{h_d}\, (A_\l + \nu) &
2\l h_d \mu - \l h_u (A_\l + 2\nu) \\ & &
\l^2 A_\l \displaystyle\frac{h_u h_d}{\mu}\, + \nu (A_\k + 4 \nu)
\ea
\right]
\eeq

\noi where $\nu = \k s$. To a good approximation, the $2\times 2$ doublet
subsector is diagonalized by the angle $\beta$ which gives the desired
light eigenstate $h$ and a heavy eigenstate $H$ with a mass
$m_H\sim m_A$ close to the MSSM-like CP odd state (the larger $m_A$,
the better this approximation). In the NMSSM, one can define $m_A^2$
as the diagonal doublet term in the CP odd $2\times2$ mass matrix after
the Goldstone mode has been dropped. At tree level, it has the same
expression as in the MSSM:
\beq\label{2.2e}
m_A^2 = \frac{2\mu B}{\sin2\beta} \ , \quad \rm{with} \quad B=A_\l + \nu \ .
\eeq

In the CP even sector of the NMSSM this is not the end of the story,
however: the light eigenstate $h$ of the $2\times 2$ doublet
subsector still mixes with the singlet state $S$, which is heavier than
$h$ by assumption. In order to maximize $m_h$, this mixing has to
vanish:
\beq\label{2.3e}
\l \left[ 2\mu - (A_\l + 2\nu)\sin2\beta \right] \sim 0 \ .
\eeq

\noi This requires either $\l \to 0$ (which minimizes the NMSSM specific
tree level contribution to $m_h$ of eq.~(\ref{1.2e})) or
\beq\label{2.4e}
A_\l \simeq \frac{2\mu}{\sin2\beta} - 2\nu\ .
\eeq

On $A_\k$ we get the following constraints: ${\cal M}_{S,33}^2$
in eq.~(\ref{2.1e})
must at least be positive, which requires essentially (since the first
term is typically relatively small) $A_\k \nu \ \gsim\ -4\nu^2$.
The CP odd mass matrix element in the singlet sector, given by
\beq\label{2.5e}
{\cal M}_{P,33}^2  =  4 \l \k h_u h_d + \l^2 A_\l \frac{h_u h_d}{\mu}\,
-3 \nu A_\k \ ,
\eeq

\noi must also be positive. Typically the last term in (\ref{2.5e})
dominates, hence we get an allowed window 
\beq\label{2.6e}
-4\nu^2\ \lsim\ A_\k \nu \ \lsim\ 0\ .
\eeq

Next, in order to maximize the NMSSM specific tree level contribution
to $m_h$ of eq.~(\ref{1.2e}), $\l$ has to be as large as possible; we
require, however, the absence of a Landau singularity for all Yukawa
couplings $\l, \k, h_t$ below the GUT scale, which leads to the
following constraints:

First, given the corresponding RG equations~\cite{nmssm}, this implies
small values for $\k$. The limit $\k \to 0$ while $\l$ remains finite
is disallowed, however, both from the stability of the potential and
the fact that the allowed window of eq.~(\ref{2.6e}) vanishes in this
limit. (Of course, stability of the potential and positivity of all
masses squared are related issues.)

Second, for small $\tanb$ the top quark Yukawa coupling becomes
large, and can run into a Landau singularity below $M_{GUT}$, or induce
a Landau singularity  below $M_{GUT}$ for $\l$. The value of
$\tanb$ that allows for maximal values of $\l$ (and maximizes the
tree level contribution to $m_h$ of eq~(\ref{1.2e})) is around 2.

In this region of $\tanb$, a larger value for the top quark pole
mass does hardly increase the upper bound on $m_h$: at fixed
$\tanb$, larger $m_{top}$ implies a larger top Yukawa coupling 
$h_t$, which implies a somewhat lower allowed value for $\l$.
Consequently a
variation of the top quark pole mass between 171.4 and 178~GeV
(which increases $m_h$ by $\sim 4.8$~GeV for $\tanb \sim 10$),
increases the maximal allowed value for $m_h$ in the NMSSM by
only $\sim 1.5$~GeV for $\tanb \sim 2$.

For large values of $\tanb$, it is obvious from eq.~(\ref{2.2e}) that
$m_A$ tends to be very large, unless $B$ is small. ($\mu$
cannot be smaller than $\sim 100$~GeV due to the lower bound on
chargino masses from LEP). Very large values of $m_A$ are unnatural,
since they require supersymmetry breaking Higgs masses of the same
order of magnitude, which aggravate the fine tuning problem -- a
situation which we want to avoid. In the MSSM, one can always chose
$B$ small enough to keep $m_A$ reasonable even at large $\tanb$.
In the NMSSM this is also possible, provided that
$\nu \simeq -A_\l$. However, one has also to minimize the
doublet-singlet mixing of eq.~(\ref{2.3e}) in order to to
maximize $m_h$. If $\l$ is not very small, eq.~(\ref{2.4e}) together with
$\nu \simeq -A_\l$ implies $\nu \!\gg\! \mu$,
which is equivalent to $\k \!\gg\! \l$. Large values of $\k$ leading to a
Landau singularity below the GUT scale, this is excluded. Thus, the only
way of minimizing the doublet-singlet mixing while keeping $m_A$
constant at large $\tanb$ is to assume $\l \to 0$, which means that
the bound on $m_h$ is the same as in the MSSM.

(In general, for large values of $\tanb$ the LEP constraints on $m_h$
imply either that $m_A$ and $|A_\l|$ assume very large values $\gsim
1$~TeV, or $\l \lsim 0.2$.)

All these considerations make it clear that a realistic upper limit on
$m_h$ in the NMSSM requires numerical methods; analytic approaches
can be misleading (and can allow for larger values of $m_h$).

Our results below are obtained with \NMHDECAY~\cite{nmhdecay3}. The
precision of the included radiative corrections to the lightest CP even
Higgs mass has already been discussed in the introduction and is
given in~\cite{nmhdecay} and~\cite{nmhdecay2}.

As discussed in the introduction, we take  universal squark and slepton
masses of 1~TeV, and trilinear squark/slepton couplings of 2.5~TeV
(near maximal mixing). For the gaugino masses we take $M_1 = 150$~GeV,
$M_2 = 300$~GeV and $M_3 = 1$~TeV. We scan over the NMSSM specific
Yukawa couplings and trilinear soft terms $\l$, $\k$, $A_{\l}$ and
$A_{\k}$ as well as the effective $\mu$ parameter, and we obtain the
regions in the NMSSM parameter space that maximize $m_h$ in agreement
with the considerations above.

\begin{figure}[t]
\begin{center}
\epsfig{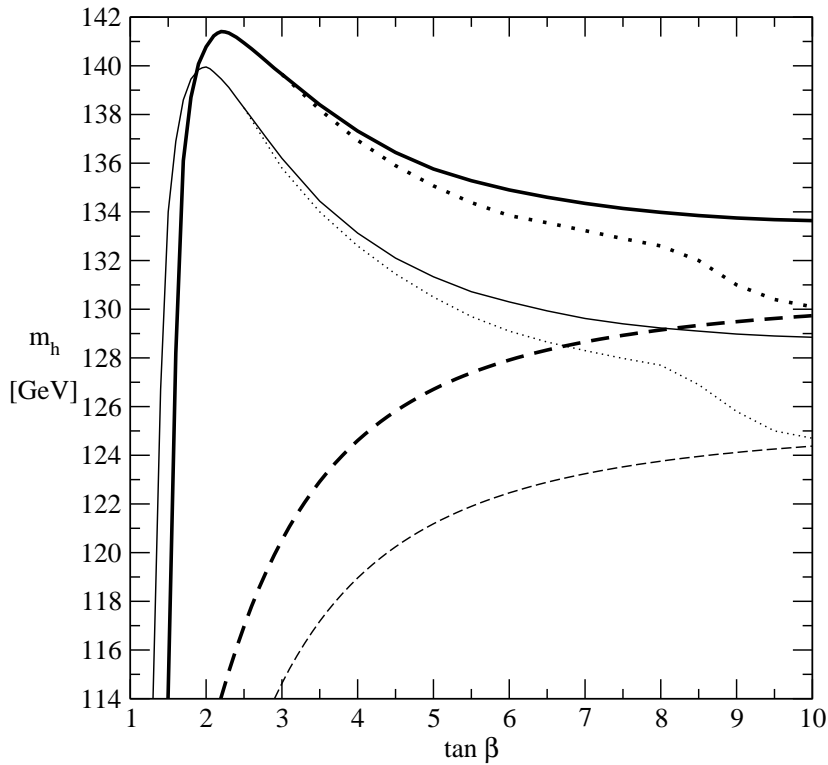}
\end{center}
\caption{Upper bound on the lightest Higgs mass in the NMSSM for
$m_{top}=178$~GeV (thick full line: $m_A$ arbitrary, thick dotted line:
$m_A = 1$~TeV) and $m_{top} = 171.4$~GeV (thin full line: $m_A$
arbitrary, thick dotted line: $m_A = 1$~TeV) and in the MSSM
(with $m_A = 1$~TeV) for
$m_{top}=178$~GeV (thick dashed line) and $m_{top} = 171.4$~GeV (thin
dashed line) as obtained with \NMHDECAY\ as a function of $\tanb$.
Squark and gluino masses are 1~TeV and $A_{top}=2.5$~TeV.}
\end{figure}

In fig.~1 we show our results  for the the upper bound on $m_h$ for
$1 < \tanb < 10$. The thick full line corresponds to $m_{top} = 
178$~GeV, the thin full line to $m_{top} = 171.4$~GeV, both without
imposing constraints on $m_A$.

With the above soft terms, the upper bound on $m_h$ in the NMSSM
is $141.4$~GeV for $m_{top}=178$~GeV. It is reached  for $\tanb\sim 2.2$,
$\l \sim .677$, $\k \sim .068$, $\mu \sim 545$~GeV, $A_\l \sim 1365$~GeV,
and $A_\k \sim 10$~GeV (strictly speaking a certain range of values for
$\k$, $\mu$, $A_\l$ and $A_\k$ gives the same result for $m_h$ for
these values of $\tanb$ and $\l$). For $m_{top}=171.4$~GeV, the upper
bound on $m_h$ is $139.9$~GeV and is obtained for $\tanb\sim 2$,
$\l \sim .703$, $\k \sim .049$, $\mu \sim 534$~GeV, $A_\l \sim 1287$~GeV
and $A_\k \sim 10$~GeV.

For $\tanb = 10$ we get $133.6$~GeV in the NMSSM for $m_{top}=178$~GeV
(resp. $128.8$~GeV for $m_{top} =  171.4$~GeV), which remains nearly
constant for larger values of $\tanb$ (a slight increase of the
contributions from the radiative corrections is compensated by a slight
decrease of the tree level term of eq.~(\ref{1.2e})).

In the same fig.~1, we show the upper bound on $m_h$ in the MSSM
limit $\l \to 0$ as obtained with \NMHDECAY\ as a thick dashed
line for $m_{top} = 178$~GeV, and as a thin dashed line for $m_{top} =
171.4$~GeV (taking $m_A= 1$~TeV). In this limit, the upper bound on
$m_h$ reaches $129.7$~GeV for $m_{top} =  178$~GeV (resp. $124.4$~GeV
for $m_{top} =  171.4$~GeV) at $\tanb = 10$, and increases by another
1~GeV for very large $\tanb = 50$. 

As noted above, large values of $\tanb$ imply large values for $m_A$ in
the NMSSM, if $\lambda$ is kept fixed. Indeed, along the full lines of
fig.~1 the value of $m_A$ increases with $\tanb$ up to several TeV. The
consequence of fixing $m_A \leq 1$~TeV is that the maximally allowed
value of $\lambda$ decreases with $\tanb$. The corresponding effect on
the upper bound of $m_h$ is shown as a thick dotted line for $m_{top} =
178$~GeV, and a thin dotted line for $m_{top} = 171.4$~GeV in fig.~1.
Now we get an upper bound of $130.1$~GeV for $m_{top}=178$~GeV (resp.
$124.7$~GeV for $m_{top} =  171.4$~GeV) at $\tanb = 10$. For larger
values of $\tanb$, the upper bound on $m_h$ remains essentially the
same as in the MSSM.

\begin{figure}[t]
\begin{center}
\epsfig{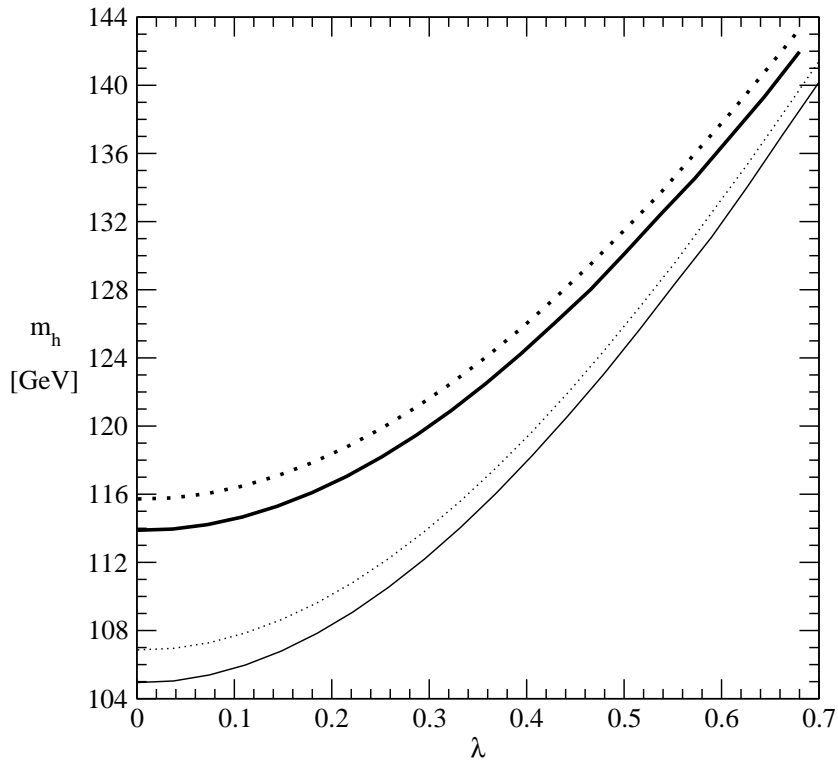}
\end{center}
\caption{Upper bound on the lightest Higgs mass in the NMSSM for
$m_{top}=178$~GeV, $\tanb=2.2$, electroweak/Yukawa corrections included
(thick full line) and omitted (thick dotted line), and $m_{top} =
171.4$~GeV, $\tanb = 2$, electroweak/Yukawa corrections included
(thin full line) and omitted (thin dotted line). Squark and gluino
masses and $A_{top}$ are as in fig.~1.}
\end{figure}

Hence, our main result is that the upper bound on $m_h$ is $\sim
12$~GeV (for $m_{top}=178$~GeV) or $\sim 16$~GeV (for
$m_{top}=171.4$~GeV) larger in the NMSSM as compared to the MSSM, and
is obtained for small $\tanb$. For very large $\tanb$, the
difference between the upper bound on $m_h$ in NMSSM and in the MSSM
vanishes, if $m_A$ is assumed to remain smaller than a few TeV.

Let us compare this bound on $m_h$ to earlier work: it is about 6~GeV
larger than the one obtained from fig.~4 in ref.~\cite{rad2b} (for the
corresponding values for $m_{top}$). Also the value of $\tanb$, where
this bound is reached, is now smaller ($\sim 2$ compared to $\sim 3$ in
ref.~\cite{rad2b}). These differences are due to the improved treatment
of radiative corrections in \NMHDECAY\, which concerns both the two
loop corrections $\sim h_t^6$ and $h_t^4\alpha_s$ (which are now
RG-improved), and the inclusion of one loop corrections (in the LLA,
keeping terms $\sim \ln(M_{Susy}^2/M_Z^2)$) proportional to the
electroweak gauge couplings and NMSSM specific Yukawa couplings
$\lambda$ and $\kappa$. The effect of the first improvement is a
considerable increase in $m_h$, whereas the effect of the
electroweak/Yukawa corrections is a slight decrease of $m_h$ by up to
$\sim 2$~GeV.

In order to clarify the latter effect and, simultaneously, the general
effect of the NMSSM specific Yukawa couplings at low $\tanb$, we show
in fig.~2 the upper bound on $m_h$ as a function of $\lambda$ at
fixed $\tanb$. Here the thick full line corresponds to $m_{top} = 
178$~GeV, $\tanb = 2.2$ and electroweak/Yukawa corrections included,
whereas the thick dotted line would be the result with these corrections
omitted. The thin full line corresponds to $m_{top} =  171.4$~GeV,
$\tanb = 2$ and electroweak/Yukawa corrections included, whereas the
thin dotted line would be the result without these corrections. One sees
the decrease in $m_h$ due to the electroweak/Yukawa corrections, which
increases the lower bound on $\l$ for small values of $\tanb$ and
$m_{top}$ due to the LEP bound on $m_h$.

As final remark we repeat, as noted at the end of the introduction,
that the mass of the lightest {\it detectable} Higgs boson could be
larger in the NMSSM than the upper bounds given here; in order to
interpret future data in the context of the NMSSM, constraints (or
positive results) must be available in the plane Higgs mass versus
Higgs couplings in order to be sensitive to a possible singlet/doublet
mixing.

\vskip 1cm
\noi {\Large{\bf Acknowledgement}}
\vskip 3mm
We like to thank P. Slavich for helpful discussions, notably for
pointing out a\break mistake in the first version of this paper. We
acknowledge support by the ANR grant\break PHYS@COL\&COS.

\end{document}